\documentclass[aps,floats,twocolumn,epsf,prl,showpacs]{revtex4-1}
\usepackage{graphicx}
\usepackage{epstopdf}

\usepackage{color}

\begin{document}

\title{Fate of the false Mott-Hubbard transition in two dimensions}

\author{T. Sch\"afer$^a$, F. Geles$^b$, D. Rost$^{c,d}$,  G. Rohringer$^a$, E. Arrigoni$^b$, K. Held$^a$, N. Bl\"umer$^c$, M. Aichhorn$^b$, A. Toschi$^a$}

\affiliation{$^a$Institute of Solid State Physics, Vienna University of Technology, 1040
Vienna, Austria}
\affiliation{$^b$ Institute of Theoretical and Computational Physics,  
  Graz University of Technology, Graz, Austria}
\affiliation{$^c$ Institute of Physics, Johannes Gutenberg University, Mainz, Germany}
\affiliation{$^d$ Graduate School Materials Science in Mainz, Johannes Gutenberg University, Mainz, Germany}

\date{ \today }

\begin{abstract}
We have studied the impact of non-local electronic correlations at all length
scales on the Mott-Hubbard metal-insulator transition in the unfrustrated
two-dimensional Hubbard model. 
Combining dynamical vertex approximation, lattice quantum Monte-Carlo and variational cluster approximation, we
demonstrate that scattering at long-range fluctuations, i.e., Slater-like paramagnons, opens a spectral gap at weak-to-intermediate coupling -- 
irrespectively of the preformation of localized or short-ranged magnetic moments.
This is the reason, why the two-dimensional Hubbard model has a paramagnetic phase which is insulating 
  at low enough temperatures for any (finite) interaction and no
  Mott-Hubbard transition is observed.
 \end{abstract}

\maketitle

\let\n=\nu \let\o =\omega \let\s=\sigma


{\sl Introduction.}
 The Mott-Hubbard metal-insulator transition (MIT) \cite{MH} 
is one of the most fundamental hallmarks of the physics of electronic correlations.
Nonetheless, astonishingly little is known exactly, even for its
simplest modeling, i.e., the
single-band Hubbard Hamiltonian \cite{Hubbard}: Exact solutions for this
model are available only in the extreme, limiting cases of one and infinite dimensions.

In one dimension (1D), the Bethe ansatz shows that there is actually
no Mott-Hubbard transition \cite{1Dbook, Kawa1989, Stanescu2001}; or, in other words, it
occurs for a vanishingly small Hubbard interaction $U$: At any $U>0$ the
1D-Hubbard model is insulating at half filling. One
dimension is, however, rather peculiar:  
While there is no antiferromagnetic ordering even at temperature
$T=0$, antiferromagnetic spin fluctuations 
are strong and long-ranged, decaying slowly, i.e.,  algebraically. Also
the (doped) metallic phase is
 not a standard Fermi liquid but a Luttinger liquid. 

For the opposite extreme, infinite dimensions, the dynamical mean  field
theory (DMFT) \cite{DMFTREV} becomes exact \cite{Dinfty}, which allows for a clear-cut  and --
to a certain extent --  almost ``idealized'' description of a pure
Mott-Hubbard MIT. In fact, since in D $\!=\!\infty$ only local correlations survive \cite{Dinfty}, the Mott-Hubbard insulator
of DMFT consists of a collection of localized (but not long-range ordered)  
magnetic moments. This way, if antiferromagnetic order is neglected
or sufficiently suppressed, DMFT describes a first-order
MIT \cite{DMFTMIT,DMFTREV}, ending with a critical endpoint.  

As an approximation, DMFT  is applicable to the more realistic
cases of the three- and two-dimensional Hubbard models. However, the DMFT
description of the MIT is the very same here, since only the
non-interacting density of states (DOS) and in particular its second moment enter.
This is a natural shortcoming of the mean-field nature of DMFT:
antiferromagnetic fluctuations have  {\sl no} effect at all
on the DMFT spectral function or self-energy above the
 antiferromagnetic ordering temperature $T_N$.

 In 3D, antiferromagnetic fluctuations reduce $T_N$  sizably compared to the DMFT (see Fig. 1), although being significant only at $T \simeq T_N$. Hence, the reliability
of the DMFT results for the spectral functions is not spoilt in 3D
except for the proximity of the antiferromagnetic transition \cite{DMFT_3D,QMC_3D,DGA_3D,DCA_3D},
whereas deviations from the DMFT entropy and susceptibilities can be
significant also at higher $T$ \cite{DCA_3D, DCA_3D_Gull}. With this background, 
it is maybe not surprising, that DMFT also yields a good description
of the MIT even for realistic material cases, such as the textbook example V$_2$O$_3$ \cite{V2O3}.

Much more intriguing, and challenging, is the 2D case,  most relevant
for high-temperature superconductivity and the rapidly emerging field
of oxide thin films and heterostructures. 
In fact, this issue has been intensely debated since the Seventies: 
On the one hand, several 
analytical and numerical 
results \cite{Castellani1979,Vekic1993,Mancini2000,Groeber2001,Avella_COP,Mancini_COP} suggested that a metallic
phase is found at weak coupling, with a MIT at a finite $U_c$. 
At the same time, calculations with the two-particle self-consistent (TPSC) approach \cite{TPSC_90s,TPSC_2000s,TPSC_2000s_2} showed a pseudogap in the perturbative regime of small $U$ \cite{comment1}.
Finally, in Anderson's view \cite{Anderson1997} the 2D physics should be considered fully nonperturbative,
similarly \cite{Stanescu2001} as in 1D, yielding a Mott gap and the 
localized physics of the 2D-Heisenberg Hamiltonian for all $U>0$.

\begin{figure}[th!] 
        \centering
                \includegraphics[width=0.48\textwidth,angle=0]{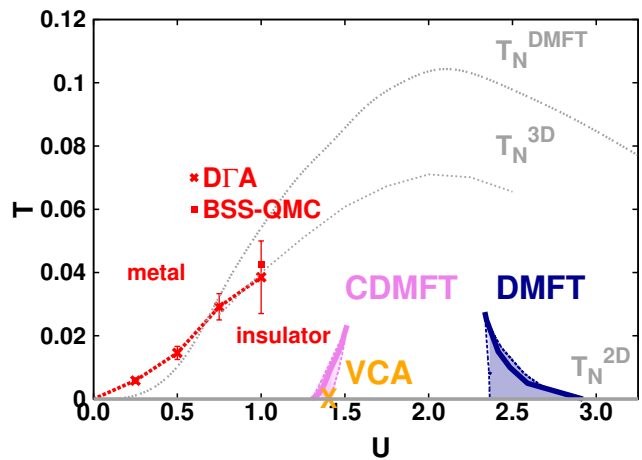}
        \caption{(color online). \label{Fig1} MIT of the Hubbard model on a square
          lattice determined by different non-perturbative
          techniques. The DMFT transition line (blue/dark \cite{bluemer_thesis})
          is shifted towards lower interaction values 
          due to short-range spatial correlations (violet/light gray line: CDMFT \cite{MIT_CDMFT}; orange cross at $T=0$: VCA). 
            This trend is
          accompanied by a simultaneous shrinking of the coexistence
          regions (hatched regions).
          The inclusion of long-range fluctuations leads  to a  vanishing 
              $U_c$ in the low-temperature regime  (crosses/red dashed line: D$\Gamma$A, red filled box BSS-QMC): Error bars mark the temperature range, where the onset of an insulating behavior on the whole Fermi surface 
        has been found, according to the electronic self-energy of D$\Gamma$A (see Fig. 3). 
 Also shown are
 the DMFT \cite{DMFT_3D} and the D$\Gamma$A 3D N\'eel temperatures (light grey dotted lines) \cite{DGA_3D} as well as the D$\Gamma$A 2D one (grey line at $T$=0) \cite{DGA2} which fulfills the Mermin-Wagner theorem \cite{Mermin}; $4t\equiv 1$ sets the energy scale.}
\end{figure}

More recently, most precise numerical studies have shown
unambiguously that the short-range
spin fluctuations do actually
reduce the critical interaction $U_c$ for the MIT in 2D compared to DMFT {\sl and} reverse
its slope, see Fig.~\ref{Fig1}. 
(Note that the DMFT insulating phase has the full entropy of free spins, i.e., 
$\ln 2$ per site,  implying the positive DMFT slope $dU_{\text{c}}/dT > 0$
of Fig.~\ref{Fig1}.) Such a 2D  picture has been established by cluster DMFT (CDMFT) \cite{MIT_CDMFT}, 
dynamical cluster approximation (DCA) \cite{DCArev,noteDCA} and
second-order dual-fermion \cite{DF} studies \cite{ThesisHartmut}, which systematically
include non-local correlations beyond DMFT. However, given the limited
cluster sizes of CDMFT and DCA calculations, only short-range 
correlations are included.

In this paper, we
revisit the MIT in 2D and the effect of antiferromagnetic
spin-fluctuations thereupon. 
To this end, we employ three methods: (i) the variational cluster
approximation (VCA) \cite{VCA} which includes short-range correlations, (ii)
the dynamical vertex approximation (D$\Gamma$A) which includes short {\sl and}
long-range correlations beyond DMFT on the same footing \cite{DGA},
and (iii) lattice quantum Monte Carlo (QMC) simulations \cite{BSS,BSS2,Wessel2014} of  unprecedented accuracy
made possible by the algorithmic progress, increased computer power and careful extrapolations (see Supplement) \cite{Bluemer2007,Rost2012}.

{\sl The phase diagram in 2D.}  Let us first summarize the  results  of our combined, comparative
studies for the half-filled Hubbard model on a square lattice with
nearest-neighbor hopping $t\equiv 1/4$ by hands of
the phase diagram Fig.\  \ref{Fig1}; all details on the 
spectra and the underlying physics of the different
regimes are presented afterwards. 

 Our VCA data for the MIT at zero temperature (orange cross in Fig.\ \ref{Fig1})
 appear consistent with the  previous CDMFT, DCA, 
 and older VCA \cite{BalzerVCA} studies,  as well
 as with second-order dual-fermion \cite{DF}  calculations \cite{ThesisHartmut}: short-range antiferromagnetic 
correlations reduce 
the critical 
$U_c$  (violet line) significantly with respect to DMFT. Moreover, the width
of the coexistence region is considerably 
reduced (see for CDMFT \cite{MIT_CDMFT} violet hatched area).
The VCA calculations performed on different clusters, however, also suggest
something more definite in this respect: At low temperatures,
the smaller the $U$, the more important becomes the effect of
longer-ranged antiferromagnetic fluctuations. 

To address this issue  in more detail,
we include  such long-range  correlations  by means of
D$\Gamma$A. Results are also compared with   lattice
Blankenbecler-Scalapino-Sugar (BSS) QMC calculations \cite{BSS}.
 The red-dashed line of Fig.\ \ref{Fig1} marks
the interaction $U_c(T)$ above which, for a given temperature $T$ 
a spectral gap is opened
because of a strong enhancement of the electronic scattering rate in the very
low-frequency regime (see below). 

These  D$\Gamma$A data, confirmed by
our extrapolated BSS-QMC data strongly suggest that
at low enough $T$ strong antiferromagnetic spin fluctuations
{\sl always} open a spectral gap,  even at arbitrarily small values of $U$ (red dashed line in Fig.\
\ref{Fig1}). Hence for $T \rightarrow 0$, $U_c \rightarrow 0$, i.e., 
{\sl no} MIT can be identified any longer
for the 2D unfrustrated Hubbard model,
 similarly as in 1D. As  we 
will elaborate in the
following, the mechanism is however rather different in this case.
By increasing $U$ the temperature of the onset of the insulating behavior is enhanced until the high-temperature crossover regime of DMFT at intermediate $U$ is reached: Here, 
the electron mobility is already suppressed  by purely local correlations.

Our results for the phase diagram indicate that the ``idealized''
physical picture of the Mott-Hubbard metal-insulator transition of
DMFT is completely overturned in 2D by strong,
spatially extended antiferromagnetic correlations.   
In the following, we will discuss explicitly the most important
aspects in terms of spatial correlations over
different length scales, and their underlying physics, by analyzing in detail the numerical
data used for determining the phase diagram in 2D. 

\begin{figure}[t!] 
        \centering
                \includegraphics[width=0.48\textwidth,angle=0]{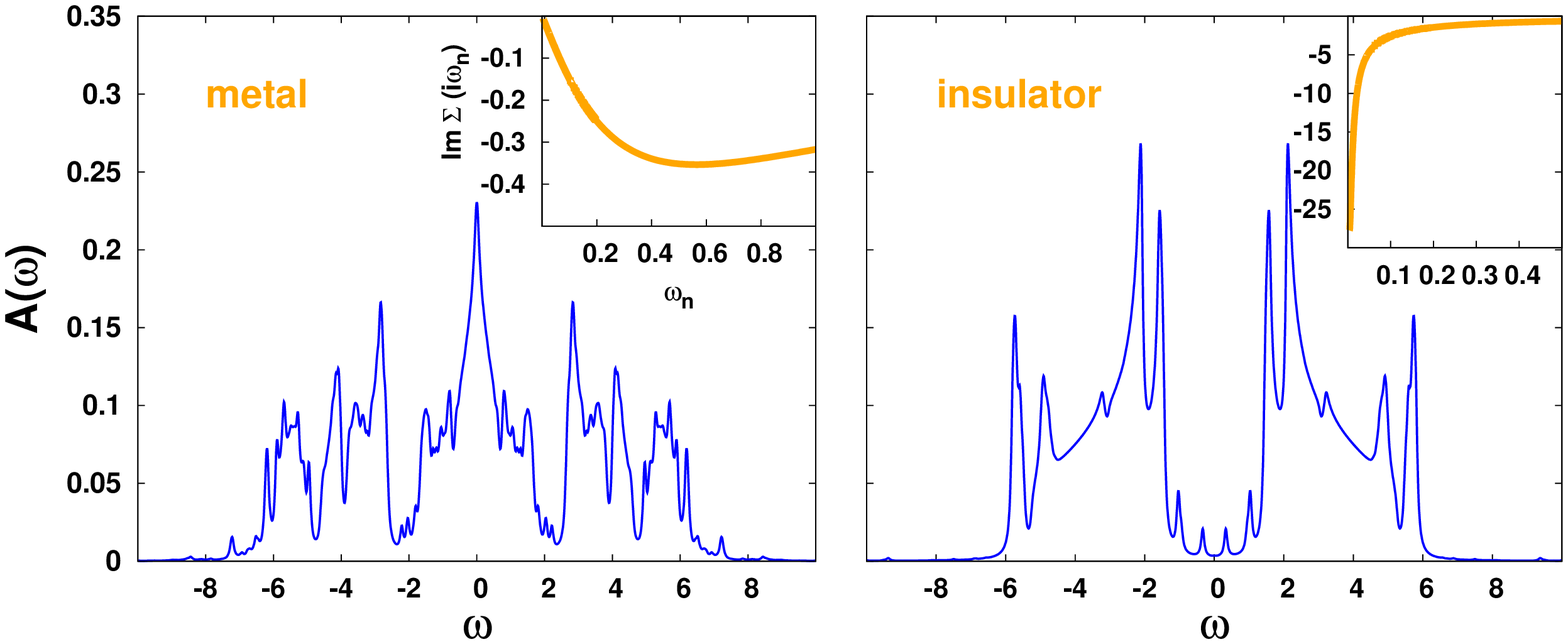}
        \caption[width=0.5\textwidth,angle=0]{(color
          online) \label{Fig2}  Local spectral function
          of the two coexisting solutions obtained in VCA  at
          the $T=0$, $U=U_c=1.4$   MIT for a $4$ site
          cluster $+$ $4$ bath sites. Left panel: 
          metallic solution; right panel: insulating solution. Inset: 
corresponding self-energies at ${\mathbf k}=(\pi,0)$.}
\end{figure}

{\sl Short-range correlations.} 
The physics of short-range correlations at $T=0$ is 
captured very well by VCA in the paramagnetic phase. 
In fact,
our results for a VCA cluster of
$N_c=4$ sites ($+ 4$ bath sites) show a clear-cut MIT at a finite $U_c
=1.4$ for $T=0$,  within the CDMFT coexistence region
 of a metallic and an insulating
solution. The local spectral function $A(\omega)$ and the self-energy $\Sigma(i\omega_n)$ at the
Fermi level of the two coexisting  solutions 
at $U=U_c= 1.4$ are reported in Fig.~\ref{Fig2}.
 The two solutions
differ qualitatively, showing a 
correlated metallic behavior with a
quasiparticle weight of $Z_{VCA}= 0.37$ at ${\bf k}=(\pi,0)$
(lower panel), and an insulating
behavior (upper panel) characterized by a divergence of Im $\Sigma(i\omega_n)$ and
a corresponding spectral gap, respectively. 
The VCA calculation of the grand potential indicates that for $U<  U_c
= 1.4$ the thermodynamically stable solution is the metallic one, while for
$U > 1.4$ the insulator is stabilized, with a level crossing at $U=U_c$. 
 Such a $U_c$ value is in fairly good agreement with CDMFT \cite{MIT_CDMFT}; it gets reduced by slightly 
 increasing the lattice size in the VCA calculations from
from  $U_c=1.4$ for  $N_c=4=2\times 2$
 to $U_c=1.325$ for $N_c=6=2\times 3$.
 This reflects
the fact 
 that  correlations of very short range
 (actually two-site in the case of $N_c=4$) are strong enough to destroy  the
 low-temperature metallic phase at intermediate coupling, but are
 less effective for lower values of the interaction.
 In fact, in the presence of a $T\!=\!0$ (magnetic) instability, a correct description of the weak-coupling regime in 2D cannot be obtained without the inclusion of correlations on {\sl  all} length scales,
as we show in the following.

\begin{figure}[t!] 
        \centering
               \includegraphics[width=0.48\textwidth,angle=0]{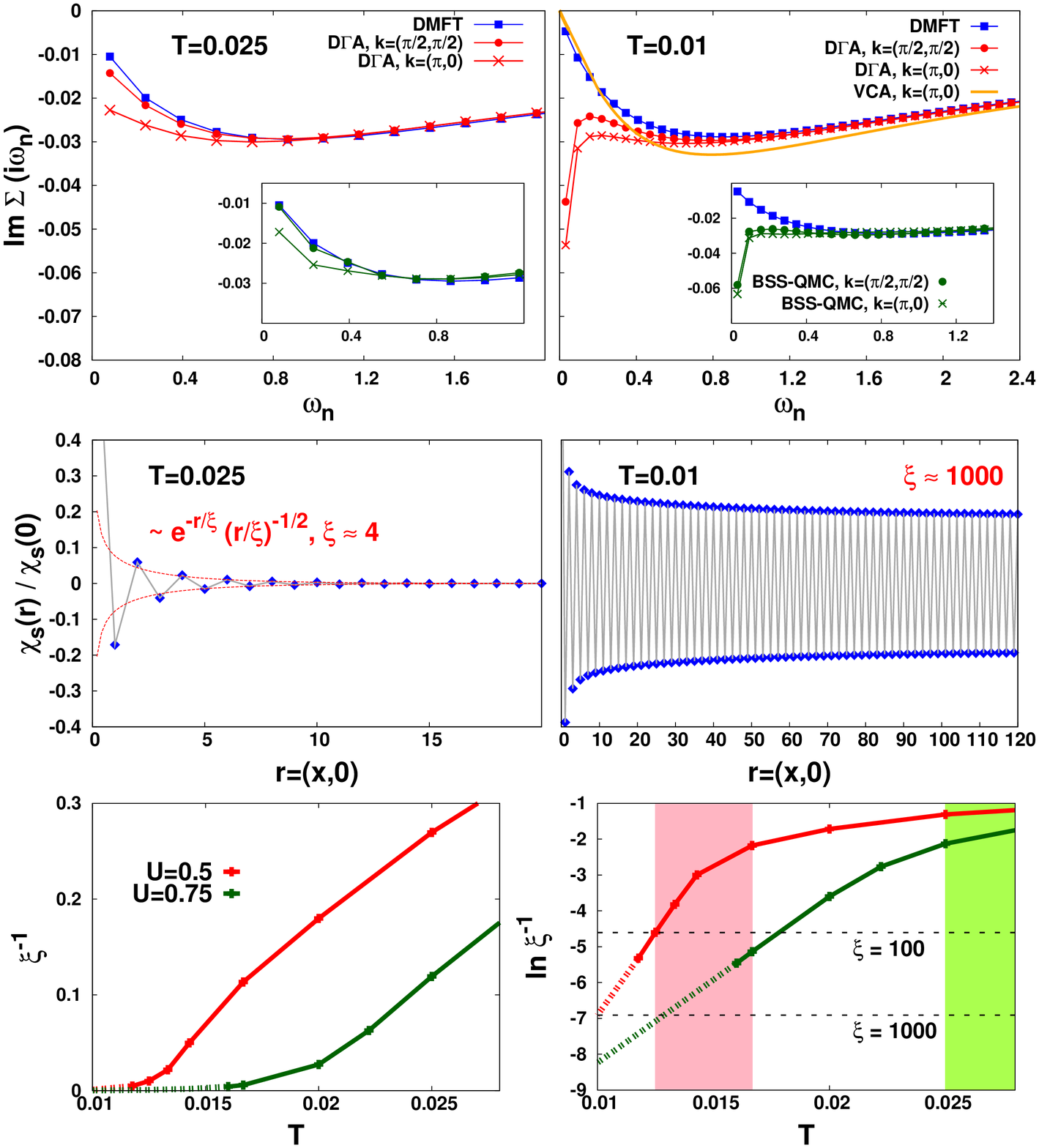}
        \caption[width=0.5\textwidth,angle=0]{(color
          online) \label{Fig3} {\sl Upper panels:} Imaginary parts of
          the self-energies for $U=0.5$, $T=0.025$ (left) and $T=0.010$
          (right), comparing DMFT
          (blue squares), D$\Gamma$A (red
          circles: ${\bf k}=(\pi/2,\pi/2)$; red crosses: ${\bf
            k}=(\pi,0)$), VCA (orange, $T=0$), and  BSS-QMC (insets, cf.\ Supplement).  
  Even for the very small interaction $U=0.5$ an insulating gap is
opened at $T\approx{0.014}$ in D$\Gamma$A as well as in BSS-QMC. {\sl  Central panels:} Real-space dependence of the D$\Gamma$A
spin correlation function $\chi_{s}({\bf r})/\chi_{s}(\vec{0})$
for the same parameters as above. Shown is the cut ${\bf r}=(x,0)$
where $x$ is given in units of the lattice spacing 
$a=1$.
 The solid line (grey, guide to the eye) interpolates between the values at
different lattice vectors (blue diamonds). By fitting (see also dashed lines in the lower panels) we obtain the
correlation lengths $\xi\approx{4}$ at  $T\!=\!0.025$ (left), while $\xi\approx{1000}$ at $T\!=\!0.010$ (right).
{\sl  Lower panels:} $T$-dependence of $\xi^{-1}$ for different interaction values. A crossover to an exponential behavior is observed at 
$T$ consistent with the onset of the insulating behavior (pink/green colored areas for $U\!=\!0.5/0.75$).
}
\end{figure}

{\sl Long-range correlations.}
We include correlations on all length scales by either
extrapolating lattice BSS-QMC results to $N_c \rightarrow \infty$ or
using D$\Gamma$A \cite{DGA}  in its
ladder version \cite{DGA2},  a
diagrammatic extension of DMFT (cf.\ \cite{DF,1PI,DMF2RG}) based on the two-particle  vertex \cite{vertex,divergence}.  
Certainly, 
both approaches have their
 limitations, either due 
to the
extrapolation procedure of the cluster results (see Supplement) or due to
the selection of the more relevant subsets of diagrams. 
Hence, cross-checking the results of these complementary
approaches, as we do here, is of utmost importance. 
In fact, the good agreement observed (upper panels of Fig.~\ref{Fig3}) validates our results and at the same time supports the physical interpretation discussed below.
The top panels of Fig. 3 show our DGA and BSS-QMC data of the imaginary part of the electronic self-energy $\Sigma({\bf k},i\omega_n)$ for
the most significant ${\bf k}-$points at the Fermi surface (i.e., the ``nodal''
point ${\bf k} =(\frac{\pi}{2},\frac{\pi}{2})$ and the ``antinodal'' point ${\bf
    k}=(\pi,0)$) as a function of 
    Matsubara
    frequencies for a rather small value of $U\!=\!0.5$ at two
different temperatures ($T\!=\!0.025$ and $T\!=\!0.010$).  Here, one can
immediately appreciate how the one-particle physics changes
even qualitatively
 when reducing $T$: At~~$\!T=\!0.025$ both D$\Gamma$A (left upper panels)
and lattice QMC (left inset)  self-energies display a Fermi-liquid behavior for all ${\bf
  k}$-points, not radically different from the DMFT results (blue
squares in Fig.\ 3). Even the  quasiparticle renormalization
$Z=(1 - \frac{\partial Im \Sigma ({\bf k} ,i\omega_n)}{\partial \omega_n}|_{\omega_n\to0})^{-1} \simeq
0.9$ 
is similar. In contrast, the
  scattering rate $\gamma$
at the Fermi surface is increased
from $\gamma_{DMFT}=- {\rm Im} \Sigma_{DMFT} ({\bf k},i0^+)\!=\!0.002,$ 
to (${\mathbf k}$-averaged) $\bar{\gamma}_{D\Gamma A} \simeq 0.014$,  with a moderate {\bf k}-differentiation\cite{Gull2010}.  
By reducing $T$,  $\gamma_{D\Gamma A}$ gets quickly enhanced on the whole Fermi surface, always displaying its largest value at ${\bf k}\!=\!(\pi,0)$. 
At $T\!=\!0.010$ the self-energy  has already changed
completely, see Fig.~3 (right): $ {\rm Im}  \Sigma({\bf k} , i\omega_n)$ acquires an evident downturn for all ${\mathbf
  k}$-points at very low frequencies. This shows that the Fermi surface is
completely destroyed at low $T$ -- even
at the nodal 
momentum ${\mathbf k}\!=\!(\pi/2,\pi/2)$. 
Such a qualitative change in the
low-frequency self-energy behavior has been exploited for defining the
(red-dashed)  line marking the destruction of the whole Fermi surface, and, hence, insulating behavior in our phase diagram. 

{\sl Physical interpretation.} Our combined numerical analysis
 not only allows us
 to make a definite statement about
the fate of the Mott-Hubbard transition in the 
2D
Hubbard model, but it
also clarifies unambiguously the physical origin of this result. 
Evidently, the shift of the border of the MIT towards  $U\!=\!0$ (Fig.~\ref{Fig1})
represents already an indication for rather extended spatial
fluctuations, emerging from the proximity to the $T\!=\!0$ long-range antiferromagnetic
order.  The important questions still to be answered: Can this intuitive picture be confirmed
 in a less heuristic and more direct way? What is the exact nature of these extended antiferromagnetic spin fluctuations?
These questions can be answered by
extending our study of the
low-$T$ weak-coupling regime to the D$\Gamma$A spin correlation
function $\chi_s({\bf r}, i\Omega_n\! = \! 0)\! = \!\int_0^\beta  d\tau \langle
S_z({\bf r}, \tau) S_z(0,0) \rangle$ in real space. Our results for $U\! =\! 0.5$ are reported in
the central panels of Fig.~\ref{Fig3}, where we show, as representative case, the
spatial decay of $\chi_s$ along the $x$-direction, normalized to its
${\bf r}=0$ value at $T \!=\!0.025$ (metal) and
$T\!=\!0.010$ (insulator):
In both cases, $\chi_s$ displays an alternating sign, which is the
typical hallmark of predominant antiferromagnetic fluctuations. The spatial
extensions of such fluctuations is quite different, however.  In fact,
the long-distance behavior of $\chi_s$ can be approximated by its asymptotic
expression $|\chi_s(r \! \rightarrow \! \infty)| \! \propto \! \sqrt{\frac{\xi}{r}} \,{e}^{-r/\xi}$\cite{Altlandbook}.
But the correlation length $\xi$
varies from $\sim \! 4$ in the metallic phase to values of $\xi\approx{1000}$ in the
low-$T$ insulating phase. A more quantitative understanding is provided by the study of the $T$-dependence of $\xi$ in D$\Gamma$A (see lowest panels of Fig. \ref{Fig3}). 
By reducing $T$, $\xi$ displays a well defined crossover to an exponential behavior, which approximately matches the onset of the low-$T$ insulating regime at weak-coupling.
This shows that the spin fluctuations responsible for the destruction of the Fermi surface at low $T$ have such a large spatial
extension, difficult to capture by (non-extrapolated) cluster
calculations\cite{White1989,LeBlanc2013}. For instance, the corresponding VCA self-energy at $T\! =\!0$ (orange
curve in Fig.~\ref{Fig3}) 
displays a very clear metallic behavior, similar to that of DMFT. 

Insight can also be gained from the potential energy. 
Our D$\Gamma$A and BSS-QMC  
results show that the destruction of the metallic state upon decreasing $T$
is accompanied by a slight reduction in potential energy,
$U \langle n_{\uparrow} n_{\downarrow} \rangle$, by about $ 1\%$
for the data of Fig. 3.
However,  this effect is
occurring in the presence of
strong and very extended ($\xi \gg 100 $) spin correlations. Therefore,
the physics cannot be really different 
from the  truly long-ranged ordered phase \cite{Wetterich2004}.
 This rules out
any particular role of prelocalization of the magnetic moments in
destroying the Fermi-liquid state, as well as the possibility of mapping
the whole low-$T$ physics onto the 2D-Heisenberg model, as proposed by Anderson \cite{Anderson1997}. Rather, the emerging physics appears more consistent to the description of the TPSC approach \cite{TPSC_2000s,TPSC_2000s_2}, at least in the weak-coupling regime, and of the low-$T$ calculations with the non-linear sigma model \cite{NLSM_2003}, as well as to the experimental estimates of $\xi$ in electron-doped  cuprates \cite{Motoyama2007}. 
In fact, the slight decrease in the potential energy is a clear
hallmark \cite{energybal, Gull2008, Gull2012} of the
Slater-like nature of the antiferromagnetic fluctuations as is the large 
 $\xi$. We can interpret this hence  as ``Slater-paramagnons''. 
The conclusive physical picture is then well defined: For all $U \!> \!0$, a
gap is opened at low enough $T$ because of the enhanced
electronic scattering with extended antiferromagnetic
paramagnons. The nature of such spin-fluctuations, reflecting the behavior
of  the $T\! =\!0$ ordered phase \cite{NLSM_2003, Lavagna2003} from which they are originating, smoothly evolves
from Slater (weak-to-intermediate coupling) to Heisenberg (strong coupling). In this respect, it is worth recalling that DCA
results \cite{Gull2008} on small clusters ($N_c\!=\! 4$) also suggest the crossover
from Slater-like  to Heisenberg-like fluctuations  for
$U$ (at least) larger than $1.25$. Though still smaller \cite{LeBlanc2013}, these interaction values are not too far away
from the regime where the crossover to Heisenberg physics is predicted to occur in the long-range ordered
phase by DMFT \cite{energybal}.

{\sl Conclusions.}
We have clarified the effects of spatial correlations on
different length scales on the MIT in the  2D
half-filled Hubbard model: for all $U>0$, at low enough (but finite) $T$, we have a paramagnetic insulator. 
This is the result of strong scattering
at extended antiferromagnetic fluctuations (paramagnons).
 The nature
of these fluctuations gradually evolves from Slater-like to Heisenberg-like, tracking an
analogous evolution for the $T\!=\!0$ antiferromagnet. This final
physical picture is quite different from both,
state-of-the art DMFT/CDMFT, which find a finite $U_c$ for the (metastable) paramagnetic phase, and the strong-coupling idea 
 of an effective low-$T$ 2D-Heisenberg model
which assumes preformed spins even at low $U$.
Instead  the 2D Hubbard model has  $U_c \!=\! 0$, and  the nature of the most relevant spin-fluctuations is
Slater-like in the whole weak-to-intermediate coupling regime.  
Let us stress that if we  frustrate the 2D square lattice away from perfect nesting, e.g., by adding a nearest-neighbor hopping,
antiferromagnetism and hence also the MIT originating from 
antiferromagnetic fluctuations is expected to shift to a finite $U_c >0$, possibly a quantum critical point.

\textit{Acknowledgments.} We thank S. Andergassen, M. Capone,  M. Fabrizio, E. Gull, O. Gunnarsson, H. Hafermann, J. Le Blanc, A. Katanin,  C. Taranto and A. Valli for discussions.
We acknowledge support from the Austrian Science Fund
(FWF) through the Doctoral School ``Building Solids for Function''  (TS, FWF project ID W1243), the
 SFB ViCoM 
(EA, FG, KH, MA, AT; FWF project ID F4103-N13) and
NAWI Graz; from
the research unit FOR 1346 of the German Research Foundation (DFG) and the graduate school GSC 266 (DR,NB). Calculations were performed on the Vienna Scientific
Cluster (VSC).

\end{document}


\title{Supplementary material to ``Fate of the false  Mott-Hubbard transition in two dimensions''}

	\author{T. Sch\"afer, F. Geles, D. Rost, G. Rohringer, E. Arrigoni, K. Held, N. Bl\"umer, M. Aichhorn, A. Toschi}


\date{\today}
  \begin{abstract}
	Direct numerical solutions of the Hubbard model in two dimensions can only be obtained for finite clusters. We show that finite-size effects are quite significant in raw BSS-QMC estimates of the self-energy, but also very regular. Consequently, the systematic errors can be reliably eliminated using finite-size extrapolations, yielding the high-precision data shown in the main paper.
 \end{abstract}
  \maketitle

The numerical results presented in the main paper have been obtained using complementary techniques with quite different characteristics. Among those, the dynamical vertex approximation ($D\Gamma A$) yields results directly in the thermodynamic limit \cite{DGA}; the variational cluster approximation (VCA), on the other hand, is good for short range correlations \cite{VCA}; and, finally, the Blankenbecler-Scalapino-Sugar (BSS) QMC calculations for the Hubbard model is applicable to clusters with a finite number $N$ of lattice sites, with $N=L^2$ for square lattices with linear extent $L$. 
In its generic formulation, the BSS-QMC algorithm introduces a further systematic bias due to a Trotter discretization of the imaginary time \cite{BSS}. In this work, we employ a multigrid approach for obtaining quasi-continuous imaginary-time Green functions without significant Trotter bias \cite{Rost2012}, which can be reliably Fourier transformed in order to compute self energies; similar strategies have proven successful in the context of DMFT studies using the Hirsch-Fye QMC algorithm \cite{Bluemer2007,Gorelik2009,*Bluemer2013}. As a result, all ``raw'' data shown in this supplement should be regarded as numerically exact for a given cluster size. 
The BSS-QMC computational effort scales as $N^3/T$ at temperature $T$, i.e. proportionally to $L^6$ at fixed $T$, which limits high-precision calculations (as we need here for determining the self-energy on the percent level) to $L\lesssim 16$. 
The properties of such finite systems will, in general, depend on the exact system size (and shape as well as boundary conditions) and may deviate drastically from the thermodynamic limit. 

We will show in the following that reliable extrapolations to the thermodynamic limit, as shown in Fig. 3 in the main paper, are still possible in the parameter range of interest based on BSS-QMC data obtained for quadratic clusters (with periodic boundary conditions) and linear extents $L=8,10,12,14,16$.

In the left column of \reff{SE_L_B25}, estimates of the self-energy $\Sigma(\mathbf{k},i\omega_n)$ at interaction $U=0.5$ and inverse temperature $\beta=100$ are shown versus Matsubara frequency $\omega_n$ for the two momenta $\mathbf{k} = (\pi,0)$ [\reffa{SE_L_B25}] and $\mathbf{k} = (\pi/2,\pi/2)$ [\reffc{SE_L_B25}], respectively; due to particle-hole symmetry the self-energy is purely imaginary at these $\mathbf{k}$ points. 
Finite-size (FS) BSS-QMC data (open symbols and broken colored lines) depend strongly on the lattice size: 
\begin{figure}[b]
	\includegraphics[width=\columnwidth]{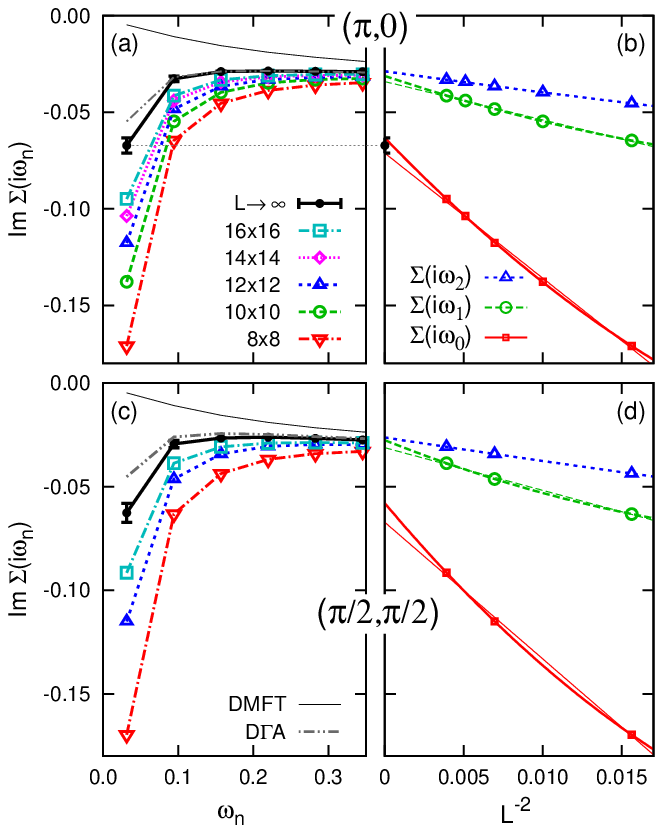}
	\caption{(Color online) Self-energy on the imaginary axis at $U = 0.5$, $\beta = 100$.
	(a) finite-size BSS-QMC data (open symbols and broken colored lines), extrapolated BSS-QMC results in the thermodynamic limit (circles and black bold solid line), and D$\Gamma$A data (grey dash-double-dotted line) versus Matsubara frequency $\omega_n$ at momentum $\mathbf{k}=(\pi,0)$; also shown are momentum-independent single-site DMFT results (thin black line). (b) finite-size BSS-QMC (symbols) data for the first 3 Matsubara frequencies versus inverse system size plus extrapolations in linear order in $L^{-2}$ (thin lines) and quadratic order (thick lines). (c) and (d) analogous analysis at $\mathbf{k}=(\pi/2,\pi/2)$.}
\label{fig:SE_L_B25}
\end{figure}
with decreasing linear extent $L$, they show increasingly insulating tendencies, i.e., larger absolute values of Im $\Sigma(i\omega_n)$ at the lowest $\omega_n$. However, as demonstrated in \reffc{SE_L_B25} for the lowest three Matsubara frequencies at $\mathbf{k} = (\pi,0)$, this bias is very systematic: Already linear extrapolations in the inverse size $L^{-2}$ (thin straight lines) yield reasonable first estimates of the the thermodynamic limit $L^{-2}\to 0$. Much better fits can be obtained in higher orders, e.g., using quadratic fits in $L^{-2}$ (thick lines); however, these become increasingly unstable (in the presence of statistical noise) at higher orders. In order to define a consistent procedure that is also stable at $\mathbf{k}=(\pi/2,\pi/2)$, where less system sizes are available (see below), we use the average of linear and quadratic extrapolation as final result, with error bars that coincide with the individual extrapolations, as illustrated by the black circle with errorbars for 
$\Sigma(i\omega_0)$ in \reffb{SE_L_B25}: 
\[
\Sigma^{\infty} = \frac{1}{2}\left( \Sigma^{\infty}_{\text{lin}} + \Sigma^{\infty}_{\text{quad}} \right) \text{, } \Delta \Sigma^{\infty} = \frac{1}{2}\left| \Sigma^{\infty}_{\text{lin}} - \Sigma^{\infty}_{\text{quad}} \right| .
\]

\begin{figure}[t]
	\includegraphics[width=\columnwidth]{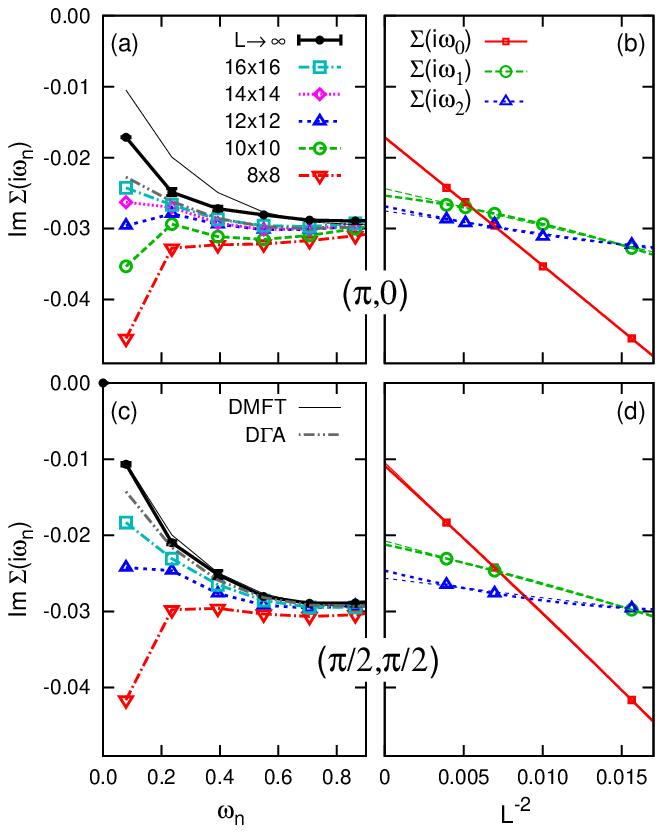}
	\caption{(Color online) Same as Fig.\ 1 but 
	at $U = 0.5$, $\beta = 40$, analogous to \reff{SE_L_B25}.}
\label{fig:SE_L_B10}
\end{figure}

The final result of this extrapolation [black circles in \reffb{SE_L_B25}] shows perfect agreement with D$\Gamma$A (grey dash-double-dotted line) at almost all Mastubara frequencies. A minor quantitative deviation is only observed at the smallest Matsubara frequency, at which the absolute value of Im $ \Sigma({\bf k,\omega)}$ is somewhat smaller in D$\Gamma$A.  

Since only lattices with linear dimensions $L=4,8,12,\dots$ contain the momentum $\mathbf{k}=(\pi/2,\pi/2)$ in the Brillouin zone (for periodic boundary conditions), we only have three system sizes available for extrapolation in this case [symbols in \reffd{SE_L_B25}]. However, the curvatures of the (here necessarily perfect, but intrinsically somewhat unstable) quadratic fits agree well with those obtained at $\mathbf{k} = (\pi,0)$, which supports their reliability. Again, the D$\Gamma$A prediction (here: a metallic self-energy with a visible momentum differentiation) agrees well with the final BSS-QMC results [black circles in \reffc{SE_L_B25}].

At the elevated temperature $T=1/40$, the finite-size bias affects the raw BSS-QMC results even more drastically, as seen in \reff{SE_L_B10}: at both $\mathbf{k}$ points, the smallest systems ($8\times 8$, red downward triangles) have clearly insulating character, while D$\Gamma$A (dash-double-dotted line) yields a metallic solution, just like (paramagnetic) DMFT (thin grey line). However, the $16\times 16$ system (squares) is already large enough to show significant metallic tendencies. Even more importantly, \reffb{SE_L_B10} and \reffd{SE_L_B10} demonstrate that the dependency of the raw BSS-QMC data on $L^{-2}$ is very regular and almost linear again (even across the FS induced metal-insulator crossover), so that the extrapolation $L^{-2}\to 0$ is still reliable, with even smaller resulting error bars than at $T=1/100$. Interestingly, the final BSS-QMC results at $\mathbf{k}=(\pi/2,\pi/2)$ [black circles in \reffb{SE_L_B10}] agree with DMFT within error bars, only at $\mathbf{k} = (\pi,0)$ nonlocal AF 
correlations induce a significantly more insulating character.